# COVID-19 AGENT BASED MODEL WITH MULTI-OBJECTIVE OPTIMIZATION FOR VACCINE DISTRIBUTION

**Jose Marie Antonio Minoza\*, Vena Pearl Bongolan and Joshua Frankie Rayo**

Department of Computer Science, College of Engineering

University of the Philippines Diliman, Quezon City, Philippines

e-mail: jminoza@up.edu.ph\*, bongolan@up.edu.ph, jbrayo@up.edu.ph

## Abstract

Now that SARS-CoV-2 (COVID-19) vaccines are developed, it is very important to plan its distribution strategy. In this paper, we formulated a multi-objective linear programming model to optimize vaccine distribution and applied it to the agent-based version of our age-stratified and quarantine-modified SEIR with non-linear incidence rates (ASQ-SEIR-NLIR) compartmental model. Simulations were performed using COVID-19 data from Quezon City and results were analyzed under various scenarios: (1) no vaccination, (2) base vaccination (prioritizing essential workers and vulnerable population), (3) prioritizing mobile workforce, (4) prioritizing elderly, and (5) prioritizing mobile workforce and elderly; in terms of (a) reducing infection rates and (b) reducing mortality incidence. After 10 simulations on distributing 500,000 vaccine courses, results show that prioritizing mobile workforce minimizes further infections by 24.14%, which is better than other scenarios. On the other hand, prioritizing the elderly yields the highest protection (439%) for the Quezon City population compared to other scenarios. This could be due to younger people, when contracted the disease, has higher chances of recovery than the elderly. Thus, this leads to reduction of mortality cases.

## Introduction

The COVID-19 pandemic is an ongoing outbreak, caused by severe acute respiratory syndrome coronavirus 2 (so-called SARS-CoV-2). The outbreak was first reported in Wuhan, China in

December 2019 [2]. Researchers around the world are working persistently to create a vaccine against COVID-19. As of December 28 2020, there were currently 15 COVID-19 vaccine candidates within Emergency Use Listing (EUL) evaluation at the World Health Organization (WHO) [16]. Now that vaccines are developed, it is very important to plan on how it will be distributed, given that the number of infections continue to rise.

Prior to vaccine development, several countries implemented lockdown measures to address the current pandemic. Additionally, health experts recommended face masks and physical distancing as part of the comprehensive strategy to suppress virus transmission. In our previous work on age-stratified and quarantine-modified SEIR with non-linear incidence rates (ASQ-SEIR-NLIR) compartmental model [6, 15, 4], these factors were considered by introducing the following parameters in the classic SEIR model (see Equations 1-4): (i) quarantine $Q(t)$, (ii) age-stratification $U$, and (iii) $\alpha$ and $\varepsilon$ to represent behavioral and disease-resistance factors that disrupt transmission.

In agent-based modeling (ABM), a system is modeled as a collection of autonomous decision-making entities called agents. Each agent individually assesses its situation and makes decisions on the basis of a set of rules. [3] The main advantage of using ABM is it could capture the emergent phenomena such as spread of the virus during interaction of agents. Second, developing ABM is flexible enough to design rules within the simulation environment. Thus, we tried to improve our existing model by transforming the ASQ-SEIR-NLIR model into ABM. It is known that during this pandemic, vaccines that will be developed will be limited. Given that COVID-19 pandemic has an incredible impact on global economic growth, there would not be enough purchasing power, especially in third world countries like the Philippines, to acquire vaccines. Thus, we need to devise a strategy for vaccination and identify what should be prioritized to equitably distribute this scarce supply. Combined with the ABM, a resource optimization model was proposed in this study to simulate possible decisions of policy makers and to help them identify appropriate strategies for their constituents.

**Methodology**

In this study, we converted the ASQ-SEIR-NLIR compartmental model [7, 16, 5] into an agent-based model to simulate people living in a common environment. We also developed a multi-objective linear programming model as a vaccine distribution priority optimization framework that aims to minimize further infections. After combining the two models, we explored different vaccination scenarios and analyzed their effects on the number of infections and the coverage of protection on the population.

**Data Sources and Area of Interest**

The primary data used in this study is the Philippines' Department of Health (DOH) COVID-19 Data Drop. Additionally, the published reports from Philippine Statistics Authority (PSA), such as The Women & Men in National Capital Region: 2018 Statistical Handbook First Edition [13], NCR Gender Factsheet [12], and 2020 RSET NCR [11] were also used for the vaccination distribution model. The area of interest for this study was Quezon City since it is considered as one of the "coronavirus hotspots" in the Philippines [9].

**COVID-19 Agent Based Model (ABM)**

The ASQ-SEIR-NLIR compartmental model is described by the following equations:

$$S' = \frac{-\beta Q(t) SI/N}{(1+\alpha S/N)(1+\varepsilon I/N)} \quad (1)$$

$$E' = \frac{\beta Q(t) SI/N}{(1+\alpha S/N)(1+\varepsilon I/N)} - \sigma UE \quad (2)$$

$$I' = \sigma UE - \gamma R \quad (3)$$

$$R' = \gamma R \quad (4)$$

The parameters of the classic SEIR model are defined by the following: β is the rate of disease transmission due to infectious population, σ is the rate of progression from exposed to infectious (reciprocal of the incubation period), and γ is the recovery rate of infectious individuals (reciprocal of the infectious period). Additionally, we introduce Q(t) as quarantine factor [16], which controls interactions between Susceptibles and Infected. *U* is age-stratified infection expectation [5] that serves as a damping factor for infections. Furthermore, we added Crowley-Martin incidence rates as nonlinear infection rates [7] α and ε, representing the inhibition effect due to susceptible and infectious population, respectively. For this model, it is assumed that all compartments of SEIR are well-mixed and interact homogeneously with each other. The coefficients β, σ and γ are considered immutable properties of the virus while α and ε are properties of the population S and I, respectively.

To capture nuances in geographic distribution, we decided to transcribe the ASQ-SEIR-NLIR into an agent-based model. [13] Here in this study, Mesa [6] as an agent-based modeling framework was utilized.

The proposed agent-based approach aims to emulate a closed society living in a shared finite environment, could be divided into states or districts and composed of Person agents with attributes corresponding to ASQ-SEIR-NLIR factors. From the classic SEIR, Removal is either via recovery with permanent immunity, or death.

In summary:

**Table 1. Agent Based Model Parameters**

| Agent Based Parameters | Compartmental Model Equivalent |
|---|---|
| Transmission Rate | β |
| Incubation Period | σ |
| Recovery Period | γ |
| People Wearing Masks, Minimum Physical Distance | α and ε |
| Natural Immunity, Persons with Pre-existing Medical Condition | α and ε |
| Maximum Minor Age Restrictions, Minimum Adult Age Restrictions | Q(t) |
| Age Infection Probability | U |

Using the Mesa framework for ABM, Agent Based Parameters shown in table above was created. These input parameters will be used for initializing Person agents' attributes (see Figure 1)

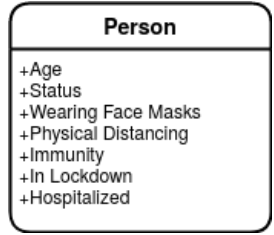

Figure 1. Person Agent

As shown in Figure 1, Person Agents have the attributes Age, Status, Wearing Face Masks, Physical Distancing, Immunity, In Lockdown and Hospitalized. Status attribute can be set by Susceptible, Exposed, Infected, Removed (*Recovered or Dead*) or Vaccinated.

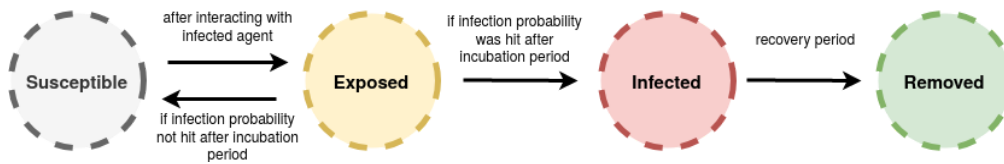

Figure 2. Agent Based Model - Status Transition

As shown in Figure 2, Susceptible, Exposed, Infected and Removed were the main status of the Person agent. In the beginning, agents will be generated using the actual age stratified population distribution data and have attributes based on input parameters. Agents wearing face masks and practicing physical distancing were primarily protected from being infected. Agents under the Maximum Minor Age Restrictions and within the Minimum Adult Age Restrictions, will not move and interact with others, having In Lockdown attribute set to true. Susceptible agents contacted with Persons having an Infected status will be under the Exposed status. During the incubation period, the age infection probability (based on Hubei incidence data_ will determine if the agent will be transferred to Infected status. Agents with Infected status could spread the virus until transmission rate probability is hit. Then, agents will be hospitalized and locked down in its current cell position. During the recovery period, an age stratified mortality rate based on the case incidence data will be calculated and will be used as probability for removed status, (recovered with permanent immunity or died). Vaccination of agents will only be implemented once, on a selected vaccination day implementation. A further assumption is 100% efficacy. Vaccines with less than 100% efficacy will have their doses divided, e.g., if a vaccine has only a 50% efficacy, the number of doses (courses) **N** will be **N/2**.

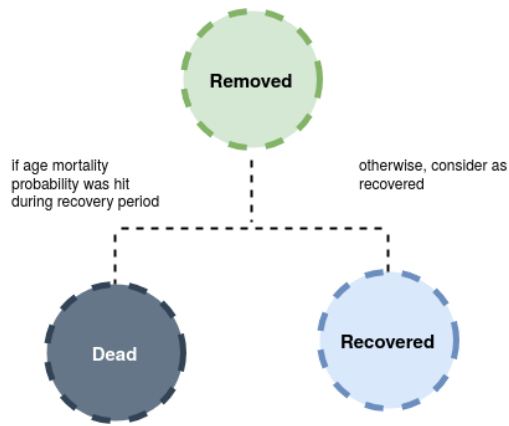

Figure 3. Removed Status

Based on the above figure, Figure 3, removed status were further categorized into Dead and Recovered status using age mortality probability. Age Infection and Mortality Probability were primarily derived from Case Incidence, Reported Recovered, Reported Died field collected on DOH Data Drop Case Information CSV file.

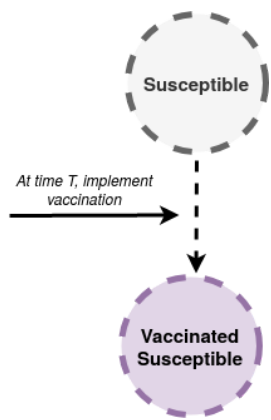

Figure 4. COVID-19 Agent Based Model with One Pulse Vaccination Strategy

Furthermore, a one-pulse vaccination strategy was implemented for integration of the vaccine distribution optimization model. In the agent-based model developed using Mesa, time T can be selected to determine the implementation day of the vaccine. Here, a portion of the susceptible population for each location are vaccinated based on the vaccine distribution optimization model.

**Multi-Objective Optimization for Vaccine Distribution**

In literature, there are various optimization models [1, 8, 15] applied in allocation of healthcare resources, including vaccine distribution. Here in this study, we formulated a linear goal optimization or multi-objective linear programming model for equitable vaccine distribution. Multi-objective linear programming deals with multiple criteria decision making problems involving more than one objective function to be optimized simultaneously, subject to linear equality and inequality constraints. Particularly for this study, our main objective function is to maximize the available vaccine doses to be distributed to minimize the further spread of the virus infection.

The formal problem addressed by the optimization model is as follows. Given the following inputs:

- **N**, a limited number of vaccine doses (courses) that can be given to remaining susceptible populations;
- **L** locations where the vaccine will be distributed;
- **A**, age stratified percentage of the population for each L locations;
- **P,** set of prioritization factors for each of the **L** locations;

The model determined the optimal allocation of available vaccines and the corresponding distribution for each location.

*Model Formulation*

Sets

    **L**            Set of Locations

    **S**            Set of Susceptible Population

| | V | Set of Vaccine Allocation of each Location |
|---|---|---|
| | P | Set of Vaccine Prioritization |
| | A | Set of Age Group |
| | C | Set of Allowed Number of Susceptibles to be Vaccinated |
| Indices | | |
| | loc | index for location |
| | s | index for susceptible population |
| | v | index for vaccine allocation |
| | p | index for vaccine prioritization |
| | a | index for age group |
| | c | index for Allowed Number of Susceptibles to be Vaccinated |
| Variables | | |
| | $S_{a,loc}$ | Susceptible Percentage Population for Location l and with Age group a |
| | $V_{loc}$ | Vaccines to be allocated for Location l |
| | $C_a$ | Allowed Number of Susceptible to be Vaccinated for Age Group a |
| | Z | Optimal Number of Vaccine to be distributed |
| Parameters | | |
| | p | Prioritization Factors (e.g. Frontliners, Mobility Workforce, Elderly) |
| | $p_{loc}$ | Prioritization Factor p for Location loc |
| | $S_{loc}$ | Age Stratified Susceptible Population for Location loc |
| | N | Limited number of vaccine doses that can be distributed |

In this optimization model, given the number of susceptible population

$$\sum_{l=1}^{L} (\sum_{a=1}^{A} S_{a,l} V_l \leq C_a) \quad (5)$$

and by limited supply N of vaccine doses

$$\sum_{loc=1}^{L} V_{loc} \leq N \quad (6)$$

our goal is to maximize the vaccine to be allocated for each location l

$$z = max \sum_{loc=1}^{L} V_{loc} \qquad (7)$$

and further optimize given with the priority factors P for each location l

$$max \sum_{p=1}^{P} ( \sum_{loc=1}^{L} p_{loc}V_{loc} \leq z) \qquad (8)$$

**Vaccine Distribution Priority Scenarios**

It is important to determine the areas and priorities to make sure that vaccines are equitably distributed. Fair Priority Model [4] was proposed as an ethical framework for vaccine global allocation. Phase I of the Fair Priority Model would allocate vaccines in order to reduce premature deaths caused by COVID-19 directly or indirectly; Phase II would aim to stem serious economic and social harms; and Phase III would seek to reduce and ultimately end community transmission. Furthermore, vulnerable groups are an important factor to consider for equitable distribution of vaccines. The definition of "most vulnerable" in the society must not be limited to adults 65 years of age or older, persons with pre-existing comorbidities, and the economically deprived. Essential workers i.e in the medical field and other most at risk should be considered. [10]

Given the PSA data for Quezon City, we used the following as Priority Factors (PF):

1. Number of Health Workers

2. Number of Public Administration and Defense

3. Number of Gainful Workers by Occupational Group

4. Persons with Functional Difficulty

5. Persons under Mobile Workforce

6. Elderly Percentage Population (60+)

In this study, frontliners for each location are primarily prioritized due to their essential roles for this pandemic. The available data concerning this are (PF1) and (PF2). Additionally, we considered Gainful Workers (PF3) as a proxy for describing the economic status of each location and Functional Difficulty (PF4) as representation of vulnerable groups.

As defined by PSA, a person with difficulty in functioning may have activity limitations, which means difficulties an individual may have in executing activities. In general, functional difficulties experienced by people may have been due to their health conditions. The concepts and definitions are based on the International Classification of Functioning, Disability, and Health (ICF) of the World Health Organization (WHO). These difficulties are the following classified into six core categories:

a. Difficulty in seeing, even if wearing eyeglasses

b. Difficulty in hearing, even if using a hearing aid

c. Difficulty in walking or climbing steps

d. Difficulty in remembering or concentrating

e. Difficulty in self-caring (bathing or dressing)

f. Difficulty in communicating

Moreover, the mobile workforce could be considered an important group especially this pandemic since various essential areas, such as public market and grocery, were limited for access and supply of necessities were becoming scarce. It is presumed that these workers could help in spreading the disease further, thus it is necessary to prioritized them for minimizing the number of infections. From the NCR Gender Factsheet [11], the following mobile workforce were considered: agriculture, construction, wholesale and retail trade, transportation and storage, and accommodation and food service activities.

It is known that the number of comorbid diseases increases with age. Therefore, prioritizing elderly could help in reducing mortality incidence (providing protection to the population). Since in the Fair Priority Model, Phase I aims to reduce premature deaths and Phase

II concerns about reducing serious economic and social deprivations, we analyzed the effect of different combinations of priority factors in terms of following objectives:

  i.  Minimizes the Number of Infections (Phase II)

  ii. Provides Highest Protection to the Population (Phase I)

For model experimentations, we consider the following scenarios: (1) no vaccination (control), (2) base vaccination, (3) prioritizing mobile workforce, (4) prioritizing elderly, and (5) prioritizing mobile workforce and elderly. In this study, we consider the priority factors 1- 4, as the base vaccination scenario, being the first concern during the implementation of vaccination distribution. In scenario 3 and 4, priority factor 5 and 6 was added to the base vaccination scenario, respectively. For scenario 5, all priority factors were considered.

To measure scenarios in terms of the objective, 10 simulation runs were conducted. In minimizing the number of infections objective, decrease in number of infected agents relative to the control scenario is computed. Meanwhile in providing highest protection to the population, the number of recovered and vaccinated agents relative to the control scenario is calculated.

## Results

**COVID-19 Agent Based Model in Quezon City**

For the simulation of the model, 10 runs were performed and the average of the results were calculated. The available vaccine to be distributed among the six districts of Quezon City was set to 500,000 courses. The December 31 2020 Case Information file from DOH Data Drop was used in calculating the initial parameters of the agent-based model. The considered Maximum Minor Age Restrictions and Minimum Adult Age Restrictions are 15 and 65, respectively. Person agents wearing masks are 93% and 53% observed social distancing. The time implementation of vaccination was set to T = 0. For Natural Immunity and Persons with Pre-existing Medical Condition parameters, the Q-SEIR-NLIR [7] $\varepsilon$ estimates were used ($\varepsilon$ = 0.1).

**Table 2. Vaccine Distribution Priority Scenarios**

| Scenarios | Minimizes Number of Infections | Provides Highest Protection |
|---|---|---|
| no vaccination (controlled) | - | - |
| base vaccination scenario | 17.24% | 434.42 |
| prioritizing mobile workforce | **24.14%** | 435.71% |
| prioritizing elderly population | 18.97% | **438.31%** |
| all priority factors | 17.24% | 437.66% |

After 10 simulation runs, results show that prioritization of mobile workforce minimizes infections 24.14% better than other scenarios . On the other hand, prioritizing the elderly, with respect to controlled scenarios, yields the highest protection (439%) for the Quezon City population, compared to vaccine scenarios (2 - 4). We note, though, that the advantage in prioritizing the elderly is small.

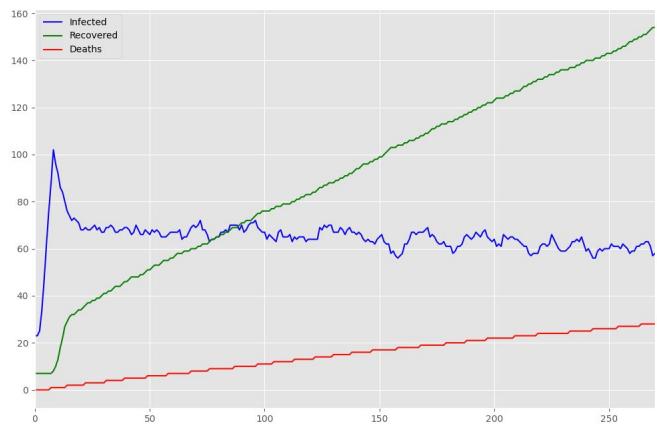

Figure 5. Plot of Agent with Infected, Recovered and Died status (Default Scenario)

As shown above, the default scenario has peaked average at 100 infected agents and after 270 iterations, has 53 agents infected.

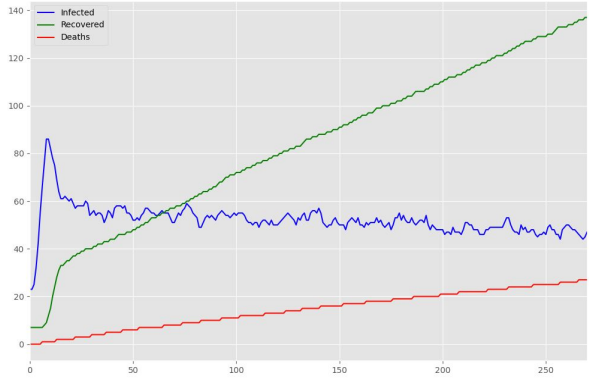

Figure 6. Plot of Agent with Infected, Recovered and Died states (Prioritizing Elderly Population)

As illustrated in Figure 6, the Prioritizing Elderly Population scenario average peaked at 87 infected agents and after 270 iterations, has only 49 agents infected.

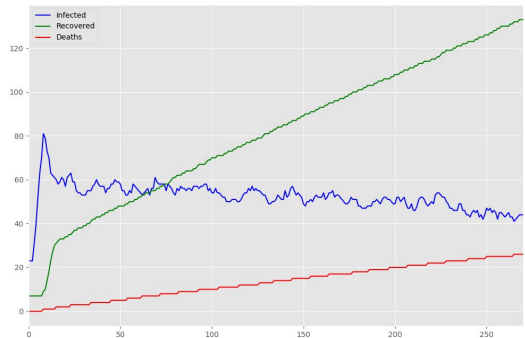

Figure 7. Plot of Agent with Infected, Recovered and Died states (Prioritizing Mobility)

On the other hand, infected agents for Prioritizing Mobility scenario average peaked at 82, complimenting with the results presented in Table 2.

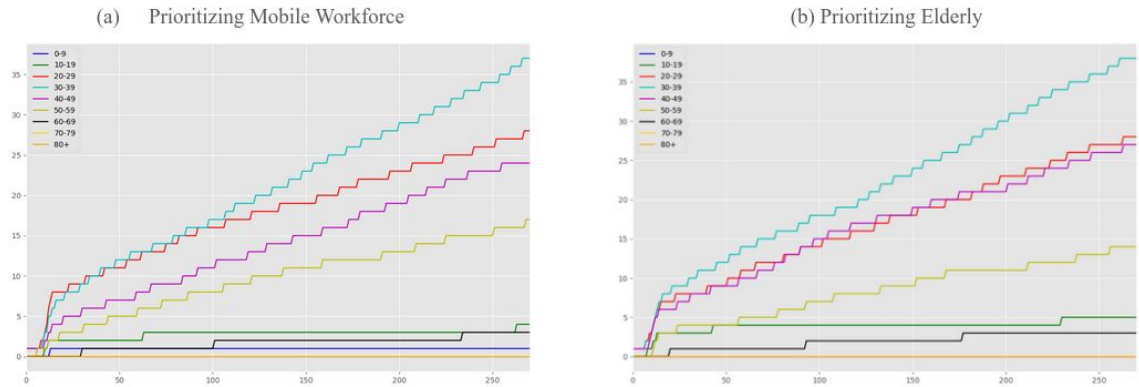

Figure 8. Comparison of Prioritizing Mobility vs Elderly (Age Stratified Recovered Agents)

Using the Data Collectors of Mesa framework, age stratified information of recovered agents was able to gather. Based on average results of 10 simulation runs, ages within 30 - 39 were effectively recovered. On the other hand, there is negligibly difference between prioritizing the mobile workforce and elderly, in terms of recovery status.

**Conclusion**

COVID-19 vaccines will soon be acquired by different countries and strategic planning on how it will be distributed is important. In summary, the ASQ-SEIR-NLIR compartmental model was able to convert into an agent-based model. Furthermore, it was improved by introducing a location-based environment and integrating it with the vaccine distribution model. An objective linear programming model as a vaccine distribution priority optimization framework was developed The model formulated was applied in Quezon City (six districts) and examined under various scenarios: (1) no vaccination scenario (control), (2) prioritizing essential workers and vulnerable population, (3) prioritizing mobile workforce scenario, (3) prioritizing elderly scenario, and (4) prioritizing mobile workforce and elderly scenario. After 10 simulation runs for distribution of one million vaccine doses or 500,000 courses, results show that aiming to minimize infections, prioritizing mobile workforce scenario works 24.14% better than other scenarios. Meanwhile, prioritizing elderly scenarios, with respect to control scenarios, yields the highest protection (439%) for the Quezon City population, compared to vaccine scenarios (2, 3 and 5).

## Declarations

**Availability of data and materials**

The datasets used and/or analyzed during the current study are available from the corresponding author on reasonable request.

**Competing interests**

The authors declare that they have no competing interests.

**Funding**

The author(s) received no specific funding for this work.

**Acknowledgements**

The authors would like to thank Prof. Roselle Leah K Rivera, Dr. Jesus Emmanuel Sevilleja, Dr. Romulo de Castro, and Dr. Salvador E. Caoili for their insightful comments during the modelling process. The authors would like to give gratitude to the efforts of undergraduate students, Karina Kylle Ang and Jimuel Celeste, for the development of ABM in the Mesa framework.